\documentclass[conference]{IEEEtran}
\usepackage{amssymb,relsize,graphicx,amsmath,psfig,cite,float,setspace,braket}
\usepackage[left=0.680in,right=0.673in,top=0.860in,bottom=1.07in]{geometry}
\usepackage{xcolor}
\usepackage{tabularx}
\usepackage{flushend}
\usepackage{algorithm}
\usepackage{algpseudocode}
\usepackage{subcaption}
\usepackage{lipsum}
\usepackage[utf8]{inputenc}
\usepackage{amsfonts, bm}
\usepackage{caption}
\usepackage{comment}
\usepackage{textcomp}

\usepackage{multirow}
\usepackage{epstopdf}
\usepackage{cite}
\usepackage{arydshln}
\usepackage{flushend}

\newcommand{\bi}{\begin{itemize}}
\newcommand{\ei}{\end{itemize}}

\newcommand{\be}{\begin{enumerate}}
\newcommand{\ee}{\end{enumerate}}
\newcommand{\bd}{\begin{description}}
\newcommand{\ed}{\end{description}}
\newcommand{\bc}{\begin{center}}
\newcommand{\ec}{\end{center}}
\newcommand{\bt}{\begin{tabbing}}
\newcommand{\et}{\end{tabbing}}
\newcommand{\bfig}{\begin{figure}}
\newcommand{\efig}{\end{figure}}
\newcommand{\beq}{\begin{equation}}
\newcommand{\beqarr}{\begin{eqnarray}}
\newcommand{\beqarrn}{\begin{eqnarray*}}
\newcommand{\eeq}{\end{equation}}
\newcommand{\eeqarr}{\end{eqnarray}}
\newcommand{\eeqarrn}{\end{eqnarray*}}
\newcommand{\bflr}{\begin{flushright}\vspace{-0.2in}}
\newcommand{\eflr}{\end{flushright}}
\newcommand{\bsub}{\begin{subequations}}
\newcommand{\esub}{\end{subequations}}
\newcommand{\barr}{\begin{array}}
\newcommand{\earr}{\end{array}}
\newcommand{\nn}{\nonumber}

\def\undb#1{\mbox{\bf{#1}}}

\def\BibTeX{{\rm B\kern-.05em{\sc i\kern-.025em b}\kern-.08em
		T\kern-.1667em\lower.7ex\hbox{E}\kern-.125emX}}

\begin{document}

\title{\huge{Performance Analysis of One- and Two-way DV-QKD\\ with MIMO FSO Communication Systems} \vspace{-0.3cm}}
%\title{\huge{Performance Analysis of One- and Two-way Protocols\\ for DV-QKD MIMO FSO Communication Systems} \vspace{-0.3cm}}

\author{ \IEEEauthorblockN{Sushil Kumar\IEEEauthorrefmark{1}, Soumya P. Dash\IEEEauthorrefmark{1}, and George C. Alexandropoulos\IEEEauthorrefmark{2}}
\IEEEauthorblockA{\IEEEauthorrefmark{1}School of Electrical and Computer Sciences, Indian Institute of Technology Bhubaneswar, Khordha, Odisha, India \\
\IEEEauthorrefmark{2}National and Kapodistrian University of Athens, Panepistimiopolis Ilissia, Athens, Greece \\
{e-mails: \{a24ec09010, spdash\}@iitbbs.ac.in, alexandg@di.uoa.gr}}
\vspace{-0.7cm}
}

{}
%{Shell \MakeLowercase{\textit{et al.}}: Bare Demo of IEEEtran.cls for Journals}

\maketitle

% \title{RIS-Assisted THz MIMO Wireless System for CV-QKD}
% \author{Sushil Kumar and Soumya~P.~Dash,~\IEEEmembership{Senior Member,~IEEE}

% \thanks{The authors are with the School of Electrical and Computer Sciences, Indian Institute of Technology Bhubaneswar, Argul, Khordha, 752050 India e-mail: (22sp06003@iitbbs.ac.in, spdash@iitbbs.ac.in).}
% }
% \maketitle

\begin{abstract}
This paper considers a multiple-input multiple-output (MIMO) wireless system wherein two legitimate users attempt to exchange secret keys over free-space optical (FSO) channels. Novel frameworks for the use of the one- and two-way discrete-variable quantum key distribution (DV-QKD) protocols, employing weak coherent pulses and decoy states, are presented. Focusing on the case where a photon-number-splitting attack is adopted by the eavesdropper and the legitimate multi-antenna receiver using threshold detection for the key extraction, novel expressions for the secret key rate and quantum bit error rate for both one- and two-way protocols are derived. The performance gain with larger MIMO configurations and the tradeoff between the performances with the one- and the two-way protocols with respect to the transmission distance of the legitimate FSO link are numerically assessed.
\end{abstract}
%%%%%%%%%%%%%%%%%%%%%%%%%%%%%%%%%%%%%%%%%%%%%%%%%%%%%%%%%%%%%%%%%%%%%%%%%%%%%%%%%%%%%%%%%%%%%%%%%%%%%%%%%%%%%%%
%%%%%%%%%%%%%%%%%%%%%%%%%%%%%%%%%%%%%%%%%%%%%%%%%%%%%%%%%%%%%%%%%%%%%%%%%%%%%%%%%%%%%%%%%%%%%%%%%%%%%%%%%%%%%%%
%%%%%%%%%%%%%%%%%%%%%%%%%%%%%%%%%%%%%%%%%%%%%%%%%%%%%%%%%%%%%%%%%%%%%%%%%%%%%%%%%%%%%%%%%%%%%%%%%%%%%%%%%%%%%%%
%%%%%%%%%%%%%%%%%%%%%%%%%%%%%%%%%%%%%%%%%%%%%%%%%%%%%%%%%%%%%%%%%%%%%%%%%%%%%%%%%%%%%%%%%%%%%%%%%%%%%%%%%%%%%%%
%%%%%%%%%%%%%%%%%%%%%%%%%%%%%%%%%%%%%%%%%%%%%%%%%%%%%%%%%%%%%%%%%%%%%%%%%%%%%%%%%%%%%%%%%%%%%%%%%%%%%%%%%%%%%%%
%%%%%%%%%%%%%%%%%%%%%%%%%%%%%%%%%%%%%%%%%%%%%%%%%%%%%%%%%%%%%%%%%%%%%%%%%%%%%%%%%%%%%%%%%%%%%%%%%%%%%%%%%%%%%%%
%%%%%%%%%%%%%%%%%%%%%%%%%%%%%%%%%%%%%%%%%%%%%%%%%%%%%%%%%%%%%%%%%%%%%%%%%%%%%%%%%%%%%%%%%%%%%%%%%%%%%%%%%%%%%%%
%%%%%%%%%%%%%%%%%%%%%%%%%%%%%%%%%%%%%%%%%%%%%%%%%%%%%%%%%%%%%%%%%%%%%%%%%%%%%%%%%%%%%%%%%%%%%%%%%%%%%%%%%%%%%%%
\vspace{-0.1cm}
\section{Introduction}
The rapid advancement towards the sixth-generation (6G) wireless communication networks aims to meet the growing demand for higher data rates and enhanced security. Non-terrestrial networks (NTNs), including satellite-based systems and high-altitude platforms, along with their seamless integration into terrestrial infrastructures, are emerging as a key driver to address these critical requirements \cite{10463684}. In recent years, free-space optical (FSO) communication systems have constituted the leading technology supporting this integration~\cite{8920091}, which is attributed to their capability to provide rapidly deployable high-capacity, interference-resistant, and license-free long-distance links. However, despite the inherent security of their narrow optical beams, FSO systems remain vulnerable to eavesdropping in dynamic environments~\cite{FSO_secrecy_phuc_sep2020_9194727}.

To address the increasing demand for unconditional security in modern communication systems, quantum key distribution (QKD) has emerged as an information-theoretic secure solution that is based on the principles of quantum mechanics. Among the various QKD schemes, discrete-variable QKD (DV-QKD) has made significant strides experimentally due to its compatibility with standard photonic components. This scheme typically utilizes single-photon or weak coherent pulse (WCP) sources to encode key information in the polarization or phase of photons \cite{bennett1984update, PracticalQKD_Scarani_sep2009}. However, since practical WCP sources emit multi-photon pulses probabilistically, they are vulnerable to photon-number-splitting (PNS) attacks. To address this issue, the decoy-state method was introduced~\cite{Hwang_aug2003}, which enables accurate estimation of the single-photon contribution, thereby ensuring secure key distribution even when dealing with imperfect sources. The authors in \cite{ma_jul2005} studied a one-way protocol for secret key exchange with the use of decoy states. A two-way DV-QKD protocol, termed LM05, was proposed in \cite{Lucamarini_apr2005}, and its performance was studied in \cite{LUCAMARINI201446} for several kinds of attacks introduced by a potential eavesdropper. Furthermore, in~\cite{SHAARI2011697}, an SKR analysis was presented for the LM05 protocol using a decoy state to enhance data security.

Although there have been ample studies on protocol designs for DV-QKD systems, they primarily lack considering the effect of several atmospheric disturbances introduced by the FSO channel, and the investigated system models are typically restricted to the single-input single-output configuration. Hence, there is a significant research gap with respect to the incorporation of multiple-input multiple-output (MIMO) configurations within DV-QKD systems that can contribute to the mitigation of the degrading effects of FSO channels. To this end, promising MIMO-based results have been lately presented for continuous-variable QKD systems \cite{sushil_11129674, sushil_RIS}. In this paper, we focus on a MIMO FSO system where two legitimate users exchange secret keys using the one- and two-way DV-QKD protocols along with decoy states, and an eavesdropper attempts to steal those keys using the PNS attack. The main contributions of this paper are summarized as follows.
\begin{itemize}
\item The transmissivity for MIMO FSO channels are derived by considering: \textit{i}) the effects of beam spreading at the transceiver; \textit{ii}) pointing error modeled using a Weibull distribution; and \textit{iii}) turbulence-induced fading modeled using a lognormal distribution.
\item Novel expressions for the secret key rate (SKR) and the quantum bit error rate (QBER) are derived for the case where the receiver deploys a threshold detector and the eavesdropper uses a PNS attack for decryption.
\end{itemize}
Numerical results demonstrate the performance gains with both protocols for MIMO systems with increasing number of antennas, showcasing also that there exists a transmission distance threshold determining where superiority between the one- and two-way protocols changes. 

% The rest of the paper is organized as follows: Section II describes the system model for the MIMO FSO DV-QKD system, including the channel model and its transmissivity. Section III discusses the transmission and reception of secret keys, as well as the effects of the eavesdropper's PNS attack on the one-way protocol. Similarly, Section IV addresses this issue in two-way protocols. Section V presents numerical results that support the analytical findings and illustrate how system performance depends on various parameters. Finally, Section VI presents the concluding remarks.

\textit{Notations}: $\textbf{A}^\dagger$ denotes the conjugate transpose of a matrix $\textbf{A}$. $J_0(\cdot)$ denotes the zero-order Bessel function of the first kind, $\jmath\triangleq\sqrt{-1}$, and $\text{diag}(\boldsymbol{a})$ constructs an $M \times M$ diagonal matrix with the elements of vector $\boldsymbol{a}$ along its principal diagonal. $\boldsymbol{0}_{N}$ represent an $N \times N$ zero matrix. $\undb{E}[ \cdot ]$ denotes the expectation operator and $\mathbb{W}$ denotes the set of non-negative integers.
%%%%%%%%%%%%%%%%%%%%%%%%%%%%%%%%%%%%%%%%%%%%%%%%%%%%%%%%%%%%%%%%%%%%%%%%%%%%%%%%%%%%%%%%%%%%%%%%%%%%%%%%
%%%%%%%%%%%%%%%%%%%%%%%%%%%%%%%%%%%%%%%%%%%%%%%%%%%%%%%%%%%%%%%%%%%%%%%%%%%%%%%%%%%%%%%%%%%%%%%%%%%%%%%%
%%%%%%%%%%%%%%%%%%%%%%%%%%%%%%%%%%%%%%%%%%%%%%%%%%%%%%%%%%%%%%%%%%%%%%%%%%%%%%%%%%%%%%%%%%%%%%%%%%%%%%%%
%%%%%%%%%%%%%%%%%%%%%%%%%%%%%%%%%%%%%%%%%%%%%%%%%%%%%%%%%%%%%%%%%%%%%%%%%%%%%%%%%%%%%%%%%%%%%%%%%%%%%%%%
\section{System and Channel Models}
The considered DV-QKD system utilizes a MIMO FSO channel between two legitimate users, Alice and Bob. Each user is equipped with $N_T$ laser sources (LSs) and $N_R$ photodetectors (PDs). The former are employed to generate WCPs, while the latter are used for threshold detection on the receiver side. The system operates either under the one-way protocol, according to which quantum states are sent from Alice to Bob, or under the two-way protocol, where Bob initiates the communication by transmitting WCPs to Alice, who then encodes their quantum information onto it and subsequently returns the encoded signal to Bob for final detection. In both cases, the goal of the legitimate FSO system is to securely exchange secret keys via the DV-QKD technique, while counteracting against eavesdropping threats realized by Eve using the PNS attack. To mitigate these attacks, decoy-state techniques are used to estimate channel parameters, detect intrusions, and ensure key confidentiality.

% In both frameworks, the objective of Alice and Bob is to establish a secure exchange of secret keys through the DV-QKD protocol, while countering the presence of a potential eavesdropper, Eve. A notable concern in this context is the PNS attack, wherein Eve exploits multi-photon pulses by isolating some photons and transmitting the remainder without inducing any significant errors. To mitigate this risk, decoy-state techniques are implemented, enabling Alice and Bob to estimate channel parameters, detect any intrusion by Eve, and ensure the confidentiality of the generated keys.

The optical beam propagation between Alice and Bob in the considered MIMO FSO system is impaired by diffraction, beam spreading, and atmospheric turbulence-induced misalignment. The overall link is modeled by an $N_R \times N_T$ complex channel gain matrix
$\mathbf{H} = \left[h_{i,j}\right]_{i=1,\ldots,N_R}^{j=1,\ldots,N_T} \in \mathbb{C}^{N_R \times N_T}$,
where the complex gain of the sub-channel from the $j$-th transmit aperture to the $i$-th receive aperture is given by~\cite{Zhao_multiplexFSO_dec2015}:
\beq
h_{i,j} = \frac{\int_{{\mathcal{D}}_i} G_j \left( r - \tilde{r} \right) \text{d}s }{\sqrt{2\pi \int_{0}^{R_0} r
\left| E_j(r) \right|^2 \text{d}r}},
\label{eq1}
\eeq
where $D_i$ is the detection area of the PD's $i$-th aperture, and $E_j(r)$ is the field distribution at the $j$-th transmitting aperture:
\beq
E_j(r) = \sqrt{\frac{2}{\pi w^2}} \exp\!\left(-\frac{r^2}{w^2}\right)
\label{eq2}
\eeq
with $w$ denoting the Gaussian beam waist at the transmitting LS (i.e., at $z = 0$, with $z$ denoting the link distance). It is noted that the received field distribution at the $i$-th aperture in~\eqref{eq1} can be obtained from the free-space propagation integral as:
\beq
G_j(r) = 2\pi \int_{0}^{\rho_{\text{max}}} \rho F_j(\rho) J_0(2\pi r \rho)e^{\jmath \sqrt{k^2-\left(2\pi\rho\right)^2} z} \text{d} \rho,
\label{eq3}
\eeq
where $k = 2\pi/\lambda$ is the wavenumber, $\rho_{\max} = \sin \left(\lambda/\pi w \right)/\lambda$, $J_0(\cdot)$ is the zeroth-order Bessel function of the first kind, and $F_j(\rho)$ denotes the spatial frequency spectrum of the transmitted field, which is computed as follows:
\beq
F_j(\rho) = 2\pi \int_0^{R_0} r \, E_j(r)\, J_0(2\pi r \rho) \text{d} r 
\label{eq4}
\eeq
with $R_0 = \sqrt{N_T} w$ being the effective transmit aperture radius.

Recall from \eqref{eq1} that the field reaching the receiver's aperture is represented by $G_j \left( r-\tilde{r}\right)$, with the term $\tilde{r}$ being introduced to model the misalignment in the FSO system. This misalignment occurs for two primary reasons: \textit{i}) pointing errors that may be caused by mechanical vibrations or imperfections in tracking; and \textit{ii}) beam wandering, which results from fluctuations in atmospheric conditions. These two factors are included in the definition of the equivalent radial standard deviation of the beam centroid displacement, $\sigma_{\tilde{r}}$. By assuming independence between these factors, we can calculate $\sigma_{\tilde{r}}$ as:
\beq
\sigma_{\tilde{r}} = \sqrt{\sigma_p^2 + \sigma_{\text{BW}}^2},
\label{eq5}
\eeq
where $\sigma_p^2 = z^2\theta_p^2$ denotes the variance caused by the pointing error with $\theta_p$ signifying the pointing jitter, and $\sigma_{\text{BW}}^2 = 0.1337\lambda^2 z^2w^{-1/3} r_c^{-5/3}$ represents the variance due to beam wandering with $ r_c = \left(0.423 \, k^2 C_n^2 \, z \right)^{-3/5}$ being the Fried parameter \cite{pirandola_FSO_mar2021, Andrews2005}. Additionally, $C_n^2$ pertains to the refractive index structure constant, which assesses the level of turbulence according to the Kolmogorov model. Typically, $C_n^2$ varies between $10^{-14}\,\text{m}^{-2/3}$ (moderate turbulence) and $10^{-17}\,\text{m}^{-2/3}$ (weak turbulence) \cite{sushil_11129674}. Finally, the displacement resulting from the misalignment factor follows a Weibull distribution having the following probability density function (p.d.f.):
\beq
f_{\tilde{r}}(v) = \frac{v}{\sigma_{\tilde{r}}^2} \exp \left(-\frac{v^2}{2\sigma_{\tilde{r}}^2}\right), \quad v \geq 0 .
\label{eq6}
\eeq

The channel matrix $\mathbf{H}$ as described in \eqref{eq1}, which considers Gaussian beam propagation, diffraction, beam spreading, beam wandering, and pointing errors, can be further expressed via its singular value decomposition (SVD) as follows:
\beq
\mathbf{H} = \mathbf{U} \boldsymbol{\Sigma} \mathbf{V}^{\dagger},
\label{eq7}
\eeq
where $\mathbf{U} \in \mathbb{C}^{N_R \times N_R}$ and $\mathbf{V} \in \mathbb{C}^{N_T \times N_T}$ are unitary matrices, and the diagonal matrix $\boldsymbol{\Sigma} \in \mathbb{R}^{N_R \times N_T}$ is represented as:
\beq
\boldsymbol{\Sigma} =
\begin{bmatrix}
\mathrm{diag}(\beta_1,\ldots,\beta_{r_H}) & 
\mathbf{0}_{r_H \times (N_T-r_H)} \\
\mathbf{0}_{(N_R-r_H)\times r_H} & 
\mathbf{0}_{(N_R-r_H)\times(N_T-r_H)}
\end{bmatrix}
\label{eq8}
\eeq
with $r_H \leq \min(N_T, N_R)$ and $\beta_i$ denoting the rank and the non-zero singular values of $\mathbf{H}$, respectively.

The effective transmissivity of each $i$-th sub-channel ($i=1,\ldots,r_H$)  also accounts for various physical impairments, including atmospheric absorption, turbulence-induced fading, and detector inefficiency. This indicates that the secret keys, when the wireless channel is used once (as in the one-way protocol), are subjected to effective transmissivity on a per $i$-th sub-channel basis, which is defined as follows:
\beq
T_i^{\text{1-way}} = \eta_d \, T_{a_i}\, T_{t_i}\, \beta_i,
\label{eq9}
\eeq
where $\eta_d$ denotes the efficiency of receiver detection, $T_{a_i}=10^{-\delta z/10}$ is the attenuation caused by atmospheric absorption with $\delta$ (dB/m) being the absorption coefficient, and $T_{t_i}$ represents turbulence-induced fading. 
Experimental research has shown that, for long-distance quantum channels, turbulence fading can be accurately modeled with a lognormal distribution \cite{Capraro_turbulance_2012}, whose p.d.f. is given per $i$-th sub-channel by:
\beqarr
f_{T_{t_i}}(u) = \frac{1}{u \sqrt{2\pi \sigma^2}}
\exp \left( -\frac{\left( \ln u 
+ 0.5\sigma^2 \right)^2}{2\sigma^2}\right),
\label{eq10}
\eeqarr
where $\sigma^2$ represents the log-irradiance variance that indicates the strength of turbulence. For weak-to-moderate turbulence, $\sigma^2$ can be determined using the Rytov approximation \cite{Andrews2005} as $\sigma^2 = e^{\xi_1+\xi_2}-1$, with $\xi_1=0.49 \chi^2/\left( 1 + 0.18 d^2\right.$ $\left. \! + 0.56 \chi^{12/5}\right)^{7/6}$, $\xi_2 = 0.51 \chi^2 \left(1 + 0.9 d^2 + 0.62 d^2 \chi^{12/5} \right)^{5/6}$,
% \beqarr
% && \!\!\!\!\!\!\!\!\sigma^2 = \exp \left(\frac{0.49 \chi^2}{\big( 1 + 0.18 d^2 + 0.56 \chi^{12/5} \big)^{7/6}} \right. \nn\\
% &&\left. \qquad \quad + \frac{0.51 \chi^2}{\big( 1 + 0.9 d^2 + 0.62 d^2 \chi^{12/5} \big)^{5/6}} \right) - 1 ,
% \label{eq11}
% \eeqarr
% with
$\chi^2 = 1.23\, C_n^2\, k^{7/6} z^{11/6}$ being the Rytov variance, and $d = a_r \sqrt{k/z}$ serving as a Fresnel number term that is influenced by the receiver aperture radius denoted by $a_r$.
%%%%%%%%%%%%%%%%%%%%%%%%%%%%%%%%%%%%%%%%%%%%%%%%%%%%%%%%%%%%%%%%%%%%%%%%%%%%%%%%%%%%%%%%%%%%%%%%%%%%%%%%%%%%%%%
%%%%%%%%%%%%%%%%%%%%%%%%%%%%%%%%%%%%%%%%%%%%%%%%%%%%%%%%%%%%%%%%%%%%%%%%%%%%%%%%%%%%%%%%%%%%%%%%%%%%%%%%%%%%%%%
%%%%%%%%%%%%%%%%%%%%%%%%%%%%%%%%%%%%%%%%%%%%%%%%%%%%%%%%%%%%%%%%%%%%%%%%%%%%%%%%%%%%%%%%%%%%%%%%%%%%%%%%%%%%%%%
%%%%%%%%%%%%%%%%%%%%%%%%%%%%%%%%%%%%%%%%%%%%%%%%%%%%%%%%%%%%%%%%%%%%%%%%%%%%%%%%%%%%%%%%%%%%%%%%%%%%%%%%%%%%%%%
\section{One-Way DV-QKD MIMO FSO System}
% This section describes the transmission and reception of secret keys via DV-QKD with the MIMO FSO system employing the one-way protocol. The performance is also studied in terms of the SKR and the QBER of the system.
%%%%%%%%%%%%%%%%%%%%%%%%%%%%%%%%%%%%%%%%%%%%%%%%%%%%%%%%%%%%%%%%%%%%%%%%%%%%%%%%%%%%%%%%%%%%%%%%%
%%%%%%%%%%%%%%%%%%%%%%%%%%%%%%%%%%%%%%%%%%%%%%%%%%%%%%%%%%%%%%%%%%%%%%%%%%%%%%%%%%%%%%%%%%%%%%%%%
\subsection{One-Way Protocol}
\begin{figure}
    \centering
    \includegraphics[width=8.5cm,height=3.8cm]{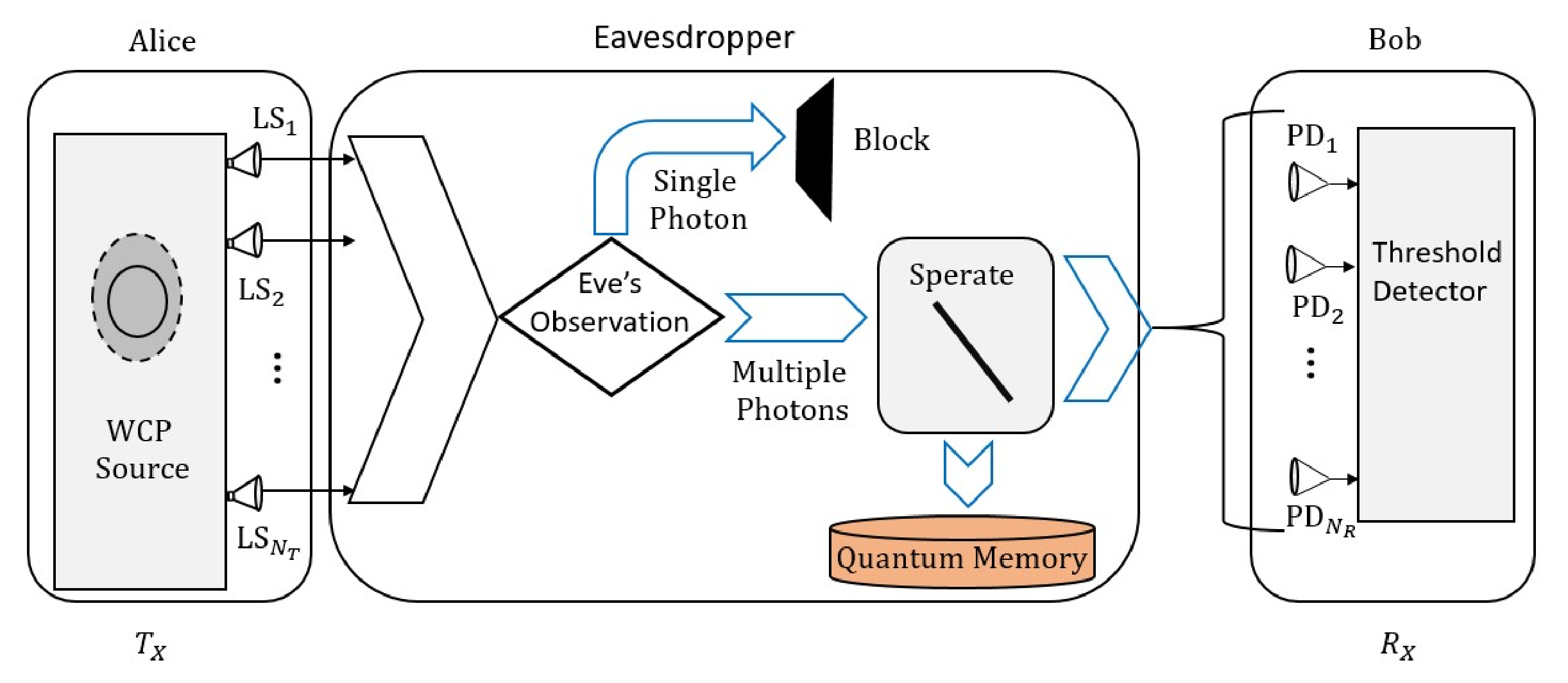}
    \caption{Model of the one-way DV-QKD MIMO FSO system.}
    \label{f1}
    \vspace{-0.3cm}
\end{figure}

% In the one-way MIMO FSO communication system, Alice employs a DV-QKD scheme to securely transmit secret keys to Bob, the legitimate receiver, while Eve, an eavesdropper, attempts to compromise the transmission by intercepting the quantum signals.
The protocol depicted in Fig. \ref{f1} includes the following three main steps: \textit{i}) Alice transmits encoded quantum decoy states; \textit{ii}) the quantum states propagate through the FSO MIMO channel, where Eve applies a PNS attack; and \textit{iii}) Bob's receiver detects and measures the incoming states using PDs.
For the transmission of the secret keys, Alice's laser system emits WCPs from each of its LSs, where the number $n_i \in \mathbb{W}$ of photons in each pulse from the $i$-th LS, $N_i$, follows a Poisson distribution having the p.d.f. expression:
\beq
\Pr \left[N_i = n_i \right]
\triangleq P_{n_i} \left( \mu_i \right)
= \frac{\mu_i^{n_i}}{n_i!}\exp\left(-\mu_i\right),
\label{eq12}
\eeq
where the mean photon number $\mu_i$ is randomly selected from $\{\mu_{s, i}, \mu_{1, i}, \mu_{2, i}\}$ $\forall\, i = 1, \ldots, N_T$, with $\mu_{s, i}$ denoting the signal's mean photon number, and $\mu_{1, i}, \mu_{2, i}$ are the mean photon numbers for the two-decoy states; it holds that $\mu_{1, i} \geq \mu_{2, i}\geq0$ and $\mu_{s, i}>\mu_{1, i}+\mu_{2, i}$. Note that this added randomness in the selection of $\mu_i$ disrupts Eve's ability to adapt to a specific photon number distribution. To this end, once $\mu_i$ has been selected to generate the WCP, Alice encodes the classical information bit onto the quantum state of the photon utilizing the BB84 protocol \cite{PracticalQKD_Scarani_sep2009}.

Following generation, the quantum signals propagate through the FSO channels, where Eve may attempt to exploit the multi-photon nature of Alice's WCPs by applying a PNS attack. More specifically, Eve can conduct a quantum non-demolition (QND) measurement to ascertain the number of photons in each incoming pulse, allowing them to gather this information without disturbing the encoded polarization state.
It is noted that, when a pulse contains a single photon, Eve cannot split it without altering its state. In this case, attempting to either forward or block the signal would disrupt Alice and Bob’s detection statistics, thus revealing the former's presence. Conversely, if the pulse contains multiple photons, Eve can split off one photon and store it in quantum memory while forwarding the remaining photons to Bob. After Alice and Bob publicly disclose their basis choices during the sifting stage, Eve can measure their stored photon in the correct basis to learn the raw key bit without introducing further errors.

At the receiving end, Bob deploys threshold detectors constituting practical implementations of DV-QKD. These detectors can effectively differentiate between a vacuum state (absence of photons) and a non-vacuum state (presence of one or more photons). However, they are incapable of ascertaining the precise number of photons in a received signal. To characterize the behavior of multi-photon signals within this model, we assume that each photon in an $n$-photon state propagates independently through the channel and Bob's detection setup for the $i$-th sub-channel, having transmissivity $T_i$ given by~(\ref{eq9}). Under this assumption, the transmittance of the $n_i$-th photon in the $n$-photon state, using a threshold detector, is given by:
\beq
T_{n,i}^{\text{1-way}} = 1 - \left(1-{T_i^{\text{1-way}}}\right)^{n_i}. %, \quad n_i \in \mathbb{W} \, .
\label{eq13}
\eeq
%%%%%%%%%%%%%%%%%%%%%%%%%%%%%%%%%%%%%%%%%%%%%%%%%%%%%%%%%%%%%%%%%%%%%%%%%%%%%%%%%%%%%%%%%%%%%%%%%
%%%%%%%%%%%%%%%%%%%%%%%%%%%%%%%%%%%%%%%%%%%%%%%%%%%%%%%%%%%%%%%%%%%%%%%%%%%%%%%%%%%%%%%%%%%%%%%%%
\subsection{Analysis of QBER and SKR}
For the protocol outlined for the one-way MIMO FSO DV-QKD system  that uses the two-decoy-state in the BB84 protocol, the SKR for the $i$-th channel is given as \cite{gllp_sep2004}:
\begin{align}
\text{SKR}_i^{\text{1-way}} \geq & 
q \bigg[ Q_{1,i}^{\text{2-decoy}}
\left(1 - H_2 \left(e_{1,i}^{\text{2-decoy}} \right)
\right) \nn \\ 
& - Q_{\mu_{s, i}} g \left(E_{\mu_{s, i}}\right) 
H_2\left(E_{\mu_{s, i}}\right) \bigg],
\label{eq14}
\end{align}
where $H_2(x) = -x \log_2(x) - (1-x)\log_2(1-x)$ is the binary Shannon entropy function, $q$ is a constant whose value is $1/2$ for the BB84 protocol, and $Q_{\mu_{s, i}}$ represents the overall gain of the $\mu_{s}$-signal state for the $i$-th channel, which is given by:
% \beqarr
% Q_{\mu_{s, i}} = \sum_{n_i=0}^\infty Q_n = Y_0 +\left(1-Y_0\right) \left(1-e^{-\mu_{s,i} T_{i}^{\text{1-way}}}\right) \, ,
% \label{eq15}
% \eeqarr
\beq
Q_{\mu_{s, i}}
\! = \! \! \sum_{n_i=0}^\infty Y_{n, i} P_{n_i} \left(\mu_{s, i}\right)
= Y_0 + \left(1-Y_0\right)
\left(1-e^{-\mu_{s,i} T_{i}^{\text{1-way}}}\right),
\label{eq15}
\eeq
where $Y_{n, i} = Y_0 + \left(1 - Y_0\right)T_{n,i}^{\text{1-way}}$ is the yield of an $n$-photon state at Bob's detection, with $Y_0$ being the background rate representing the characteristics of the photon detector at Bob's end. Moreover, $E_{\mu_{s, i}}$ is the QBER corresponding to the overall signal gain $Q_{\mu_{s, i}}$, whish is expressed as follows:
\beq
 E_{\mu_{s, i}} = \frac{\sum_{n_i=0}^\infty e_{n,i}\, Q_{n,i}}{\sum_{n_i=0}^\infty Q_{n,i}} 
=\frac{e_0Y_0 + e_{\text{det}}\left(1-e^{-\mu_{s, i}T_i^{\text{1-way}}}\right)}{Q_{\mu_{s, i}}},
\label{eq16}
\eeq
where $e_{n,i}=\left(e_0Y_0+e_{\text{det}}T_{n,i}^{\text{1-way}}\right)/Y_{n,i}$, $e_0$ is the background error rate, and $e_{\text{det}}$ signifies the likelihood of a photon striking the incorrect detector, which also reflects the alignment and stability of the optical system. Furthermore, $g \left( E_{\mu_{s, i}} \right)$ represents the bidirectional error-correction efficiency as a function of the error rate, $Q_{1,i}^{\text{2-decoy}}$ denotes the gain of a single photon in the pulse, and $e_{1,i}^{\text{2-decoy}}$ is the upper bound on the associated QBER. To compute $Q_{1,i}^{\text{2-decoy}}$, we use \eqref{eq15} and calculate:
%and $e_{1_i}^{\text{2-decoy}}$
\begin{align}
Q_{\mu_{1,i}}e^{\mu_{1,i}}
& \! \! - Q_{\mu_{2,i}}e^{\mu_{2,i}}
\! = \! Y_{1,i} \! \left(\mu_{1,i} \! - \! \mu_{2,i}\right)
\! + \! \! \! \sum_{n_i=2}^{\infty} \! \!
\frac{Y_{n,i} \! \left(\mu_{1,i}^{n_i}
\! - \! \mu_{2,i}^{n_i} \right)}{n_i!} \nn \\
& \stackrel{(a)}{\leq}
Y_{1,i} \left(\mu_{1,i} - \mu_{2,i}\right)
+ \frac{\left(\mu_{1,i}^2-\mu_{2,i}^2\right)}{\mu_{s,i}^2}
\sum_{n_i=2}^{\infty}\frac{Y_{n,i} \mu_{s,i}^{n_i}}{n_i!}
\nn\\
&\stackrel{(b)}{\leq}
Y_{1,i} \left(\mu_{1,i} - \mu_{2,i} \right)
+ \frac{\left(\mu_{1,i}^2-\mu_{2,i}^2\right)}{\mu_{s,i}^2} 
\nn \\
& \qquad \times \left(Q_{\mu_{s,i}}e^{\mu_{s,i}}
- Y_0^L - Y_{1,i}\mu_{s,i} \right).
\label{eq17}
\end{align}
Given the conditions $0 < a + b < 1$ and $n \geq 2$, step $(a)$ utilizes $a^n - b^n \leq a^2 - b^2$,  and step $(b)$ follows after some algebraic simplifications using \eqref{eq15}. Further simplifications yield the bound on $Y_{1,i}$, denoted by $Y_{1,i}^{\text{2-decoy}}$, as follows:
\begin{align}
Y_{1,i} \geq Y_{1,i}^{\text{2-decoy}}
= & \frac{\left(\mu_{s,i}- \mu_{1,i} - \mu_{2,i}\right)^{-1}}
{\mu_{s,i} \left(\mu_{1,i} - \mu_{2,i}\right)} \nn \\
& \times \left( \mu_{s,i}^2 \left( Q_{\mu_{1,i}}e^{\mu_{1,i}}-Q_{\mu_{2,i}}e^{\mu_{2,i}} \right) \right. \nn \\
& \left. - \left(\mu_{1,i}^2 - \mu_{2,i}^2 \right) 
\left(Q_{\mu_{s,i}} e^{\mu_{s,i}} - Y_0^{L}\right) \right), 
\label{eq18}
\end{align}
where $Y_0^{L} = \max \left\{\frac{\mu_{1,i} Q_{\mu_{2,i}} e^{\mu_{2,i}} - \mu_{2,i} Q_{\mu_{1,i}} e^{\mu_{1,i}}} {\mu_{1,i}-\mu_{2,i}}, \; 0\right\}$ and $Q_{1,i}^{\text{2-decoy}} = Y_{1,i}^{\text{2-decoy}} \mu_{s,i} e^{-\mu_{s,i}}$.

%%%%%%%%%
\begin{figure}[!t]
    \centering
    \includegraphics[width=8.5cm,height=5cm]{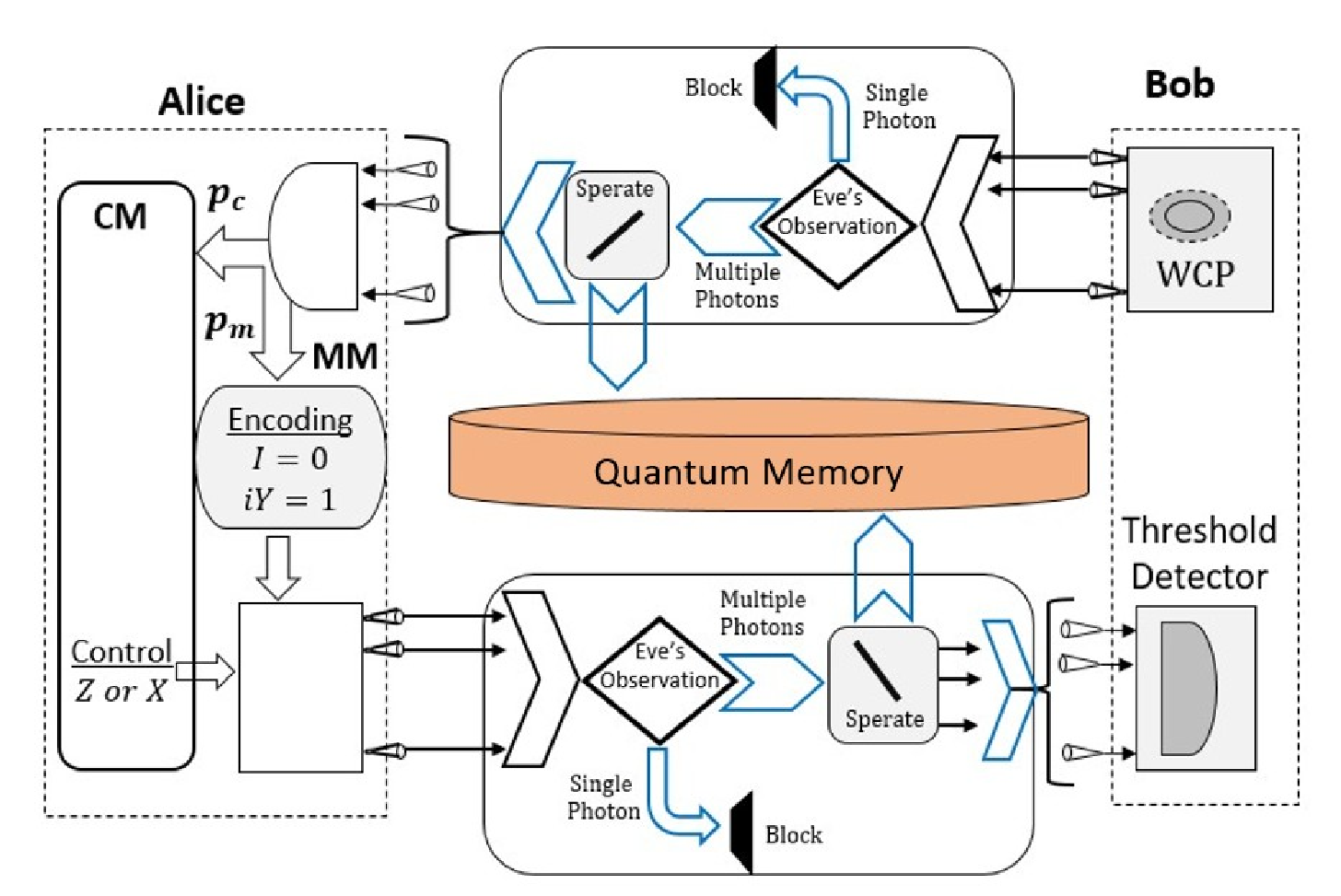}
    \caption{Model of the two-way MIMO FSO DV-QKD system.}
    \label{f2}
    \vspace{-0.5cm}
\end{figure}
%%%%%%%%%
% \beqarr
% E_{\mu_{1,i}} \! \! = \!
% \frac{e^{-\mu_{1,i}}}{Q_{\mu_{1,i}}}
% \! \left(e_0Y_0 +e_{1,i}\mu_{1,i}Y_{1,i}
% + \! \! \sum_{n_i=2}^{\infty}
% \frac{e_{n,i} Y_{n,i}\mu_{1,i}^{n_i}}{n_i!}\right) \! ,
% \label{eq17}
% \eeqarr
% and
% \beqarr
% E_{\mu_{2,i}} \! \! = \!
% \frac{e^{-\mu_{2,i}}}{Q_{\mu_{2,i}}}
% \! \left(e_0Y_0 +e_{1,i}\mu_{2,i}Y_{1,i}
% + \! \! \sum_{n_i=2}^{\infty}
% \frac{e_{n,i} Y_{n,i}\mu_{2,i}^{n_i}}{n_i!} \right) \! .
% \label{eq18}
% \eeqarr
Similarly, utilizing the expressions for $E_{\mu_{1,i}}$ and $E_{\mu_{2,i}}$ in \eqref{eq16} along with the condition $\mu_{1, i} \geq \mu_{2, i}\geq0$, we obtain the following upper-bound on the single-photon error:
\beq
e_{1,i} \leq e_{1,i}^{\text{2-decoy}}
= \frac{E_{\mu_{1,i}} Q_{\mu_{1,i}} e^{\mu_{1,i}}
- E_{\mu_{2,i}} Q_{\mu_{2,i}} e^{\mu_{2,i}}}
{(\mu_{1,i} - \mu_{2,i})\, Y_{1,i}^{\text{2-decoy}}} \, .
\label{eq19}
\eeq
Owing to the use of SVD, the overall SKR of the considered MIMO FSO system employing the one-way DV-QKD protocol is obtained as in (\ref{eq20}) at the top of the next page. 
\begin{figure*}[!t]
\beq
\text{SKR}_{\text{MIMO}}^{\text{1-way}}
= \sum_{i=1}^{{r_H}} \undb{E}_{T_i}
\left[ \text{SKR}_{i}^{\text{1-way}} \right]
= \sum_{i=1}^{{r_H}} \undb{E}_{T_i} \left[ q\left(- Q_{\mu_{s, i}} g\left(E_{\mu_{s, i}}\right) H_2\left(E_{\mu_{s, i}} \right)+ Q_{1,i}^{\text{2-decoy}}\left(1 - H_2\left(e_{1,i}^{\text{2-decoy}}\right)\right) \right)\right]
\label{eq20}
\eeq
\noindent\rule{\textwidth}{.5pt}
\vspace{-0.8cm}
\end{figure*}
In addition, the QBER for single photon pulses in the considered system setup under the BB84 protocol can be calculated as the weighted average of the QBERs $e_{1,i}^{\text{2-decoy}}$ $\forall i$, yielding:
\beq
\text{QBER}_{\text{MIMO}}^{\text{1-way}}  
= \frac{\sum_{i=1}^{{r_H}} e_{1,i}^{\text{2-decoy}}\,Q_{1,i}^{\text{2-decoy}}}
{\sum_{i=1}^{{r_H}} Q_{1,i}^{\text{2-decoy}}} \, .
\label{eq21}
\eeq
%%%%%%%%%%%%%%%%%%%%%%%%%%%%%%%%%%%%%%%%%%%%%%%%%%%%%%%%%%%%%%%%%%%%%%%%%%%%%%%%%%%%%%%%%%%%%%%%%
%%%%%%%%%%%%%%%%%%%%%%%%%%%%%%%%%%%%%%%%%%%%%%%%%%%%%%%%%%%%%%%%%%%%%%%%%%%%%%%%%%%%%%%%%%%%%%%%%
%%%%%%%%%%%%%%%%%%%%%%%%%%%%%%%%%%%%%%%%%%%%%%%%%%%%%%%%%%%%%%%%%%%%%%%%%%%%%%%%%%%%%%%%%%%%%%%%%
%%%%%%%%%%%%%%%%%%%%%%%%%%%%%%%%%%%%%%%%%%%%%%%%%%%%%%%%%%%%%%%%%%%%%%%%%%%%%%%%%%%%%%%%%%%%%%%%%
\section{Two-Way DV-QKD MIMO FSO System}
% This section describes the transmission and reception of secret keys via DV-QKD using the MIMO FSO system with a two-way protocol. The performance is again studied in terms of the SKR and the QBER of the system.
%%%%%%%%%%%%%%%%%%%%%%%%%%%%%%%%%%%%%%%%%%%%%%%%%%%%%%%%%%%%%%%%%%%%%%%%%%%%%%%%%%%%%%%%%%%%%%%%%
%%%%%%%%%%%%%%%%%%%%%%%%%%%%%%%%%%%%%%%%%%%%%%%%%%%%%%%%%%%%%%%%%%%%%%%%%%%%%%%%%%%%%%%%%%%%%%%%%
\subsection{Two-Way protocol}
In the two-way MIMO FSO DV-QKD protocol, as depicted in Fig.~\ref{f2}, Bob initiates the protocol for exchanging secure keys by generating WCP through his $N_R$-LS-based transmit aperture. Each WCP follows a Poissonian photon number distribution with mean photon number $\mu_i \in \{\mu_{s, i}, \mu_{1, i}, \mu_{2, i} \}$, where $\mu_i$ is randomly selected to apply the decoy-state method. To this end, Bob encodes the classical information bit using the LM05 protocol and transmits the prepared qubit to Alice.
Upon receiving the qubits via the MIMO FSO channels, Alice can perform either the {\em message modes (MM)} action or the {\em control modes (CM)} action, associated with probabilities $p_m$ and $p_c \neq0$, respectively, with $p_m+p_c=1$.
In the MM action, Alice encodes their information into the received signal by using an identity operation $\undb{I}$ to represent $0$ and a spin flip operation $\jmath \undb{Y}$ to represent $1$. In CM, Alice performs a projective measurement on the received signal using either the Pauli $\undb{Z}$ or the Pauli $\undb{X}$ eigenstates, with this choice made randomly \cite{LUCAMARINI201446,beaudry_2013}. We consider that Alice chooses the MM action as it enables key generation with higher efficiency and throughput, reduces communication overhead, and maintains security by requiring fewer monitoring rounds for eavesdropping detection, as compared to the CM action. Following this, the qubit is sent back to Bob, who performs a threshold detector on the received signal. Thus, the overall round-trip transmissivity (Bob-Alice-Bob) for the $i$-th sub-channel can be expressed as:
\beq
T_i^{\text{2-way}} = \eta_d \, p_m \, T_{a_i} \, T_{b_i} \, T_{t_i}^2\, \beta_i^2,
\label{eq22}
\eeq
where $\eta_d$, $T_{a_i}$, $T_{t_i}$, and $\beta_i$ are given in \eqref{eq8}, and $T_{b_i}=10^{-\delta z/10}$ denotes the atmospheric attenuation of the Alice-Bob path.

During the latter process, Eve performs a PNS attack on the communication channel in both phases of the secret key exchange. To ensure the feasibility of this scenario, it is considered that Bob generates $n_i \geq 3$ photons.
%%%%%%%%%%%%%%%%%%
\setcounter{equation}{24}
\begin{figure*}[!t]
\begin{align}
\tilde{e}_{1,i} & = \frac{\left(E_{\mu_{1,i}} Q_{\mu_{1,i}} e^{\mu_{1,i}} - E_{\mu_{2,i}} Q_{\mu_{2,i}} e^{\mu_{2,i}}\right) \left(\mu_{s,i}^2 - \mu_{2,i}^2\right) 
- \left(E_{\mu_{s,i}} Q_{\mu_{s,i}} e^{\mu_{s,i}} - E_{\mu_{2,i}} Q_{\mu_{2,i}} e^{\mu_{2,i}}\right) \left(\mu_{1,i}^2 - \mu_{2,i}^2\right)}
{Y^{L}_{1,i}\left(\mu_{s,i}-\mu_{1,i}\right) \left(\mu_{s,i}-\mu_{2,i}\right) \left(\mu_{1,i}-\mu_{2,i}\right)} \, , \nn \\
%%%%%%%%%%%%%%
\tilde{e}_{2,i} & = -\frac{2\left(\left(E_{\mu_{1,i}} Q_{\mu_{1,i}} e^{\mu_{1,i}} - E_{\mu_{2,i}} Q_{\mu_{2,i}} e^{\mu_{2,i}}\right) \left(\mu_{s,i}-\mu_{2,i}\right) 
- \left(E_{\mu_{s,i}} Q_{\mu_{s,i}} e^{\mu_{s,i}} - E_{\mu_{2,i}} Q_{\mu_{2,i}} e^{\mu_{2,i}}\right) \left(\mu_{1,i} - \mu_{2,i}\right)\right)}
{Y^{L}_{2,i}\left(\mu_{s,i}-\mu_{1,i}\right) \left(\mu_{s,i}-\mu_{2,i}\right) \left(\mu_{1,i}-\mu_{2,i}\right)}
\label{eq25}
\end{align}
\noindent\rule{\textwidth}{.5pt}
\vspace{-0.8cm}
\end{figure*}
%%%%%%%%%%%%%%%%
%%%%%%%%%%%%%%%
\begin{figure*}
\begin{align}
% Y^{L}_{1,i} & = \frac{\mu_{s,i}}{\left(\mu_{1,i} - \mu_{2,i}\right) \left(\mu_{s,i}-\mu_{1,i} - \mu_{2,i}\right)}
% \left[Q_{\mu_{1,i}} e^{\mu_{1,i}} - Q_{\mu_{2,i}} e^{\mu_{2,i}}- \frac{\mu_{1,i}^2 - \mu_{2,i}^2}{\mu_{s,i}^2} \left(Q_{\mu_{s,i}} e^{\mu_{s,i}} - Y^{L}_0\right)
% \right] \nn \\
Y^{L}_{1,i} & = \left(\mu_{s,i} \left(\mu_{1,i} - \mu_{2,i}\right) \left(\mu_{s,i}-\mu_{1,i} - \mu_{2,i}\right) \right)^{-1}
\left( \mu_{s,i}^2 \left( Q_{\mu_{1,i}} e^{\mu_{1,i}}
- Q_{\mu_{2,i}} e^{\mu_{2,i}} \right)
- \left(\mu_{1,i}^2 - \mu_{2,i}^2 \right)
\left(Q_{\mu_{s,i}} e^{\mu_{s,i}} - Y^{L}_0\right)
\right) \, , \nn \\
%%%%%%%%%%%%%%%
Y^{L}_{2,i} & = \frac{2\mu_{s,i}}{\mu_{s,i}\left(\mu_{1,i}^2 - \mu_{2,i}^2\right) - \left(\mu_{1,i}^3 - \mu_{2,i}^3\right)} 
\left[Q_{\mu_{1,i}} e^{\mu_{1,i}} - Q_{\mu_{2,i}} e^{\mu_{2,i}}
- \left( Y^{U}_{1,i} \left( \frac{\mu_{s,i}^2 \left(\mu_{1,i} - \mu_{2,i}\right) - \left(\mu_{1,i}^3-\mu_{2,i}^3 \right)}{\mu_{s,i}^2} \right) \right. \right. \nn \\
& \hspace{9.5cm} \left. \left.
+ \frac{\mu_{1,i}^3 - \mu_{2,i}^3}{\mu_{s,i}^3}\left(Q_{\mu_{s,i}} e^{\mu_{s,i}}- Y^{L}_0\right)\right)\right]
\label{eq26}
\end{align}
\noindent\rule{\textwidth}{.5pt}
\vspace{-0.8cm}
\end{figure*}
%%%%%%%%%%%%%%%%%
%%%%%%%%%%%%%%%
\setcounter{equation}{27}
\begin{figure*}[!t]
\beqarr
\text{SKR}_{\text{MIMO}}^{\text{2-way}}
= \sum_{i=1}^{{r_H}} \undb{E}_{T_i}
\left[ \text{SKR}_{i}^{\text{2-way}} \right]
= \sum_{i=1}^{{r_H}} \undb{E}_{T_i} \left[q\left(-\tilde{Q}_{\mu_s,i} f\left(\tilde{E}_{\mu_s,i}\right) H_2\left(\tilde{E}_{\mu_s,i}\right) + \sum_{n=1}^{2} Q_{n,i}^{L} \left(1 - G\left(\tilde{e}_{n,i}\right) \right)\right)\right]
\label{eq28}
\eeqarr
\noindent\rule{\textwidth}{.5pt}
\vspace{-0.8cm}
\end{figure*}
%%%%%%%%%%%%%%%%%
%%%%%%%%%%%%%%%%%%%%%%%%%%%%%%%%%%%%%%%%%%%%%%%%%%%%%%%%%%%%%%%%%%%%%%%%%%%%%%%%%%%%%%%%%%%%%%%%%
%%%%%%%%%%%%%%%%%%%%%%%%%%%%%%%%%%%%%%%%%%%%%%%%%%%%%%%%%%%%%%%%%%%%%%%%%%%%%%%%%%%%%%%%%%%%%%%%%
\subsection{Analysis of QBER and SKR}
The attainable SKR of the $i$-th sub-channel for the two-way MIMO FSO DV-QKD system utilizing the two-decoy-state in the LM05 protocol is given as follows~\cite{SHAARI2011697}:
\setcounter{equation}{21}
\begin{align}
\text{SKR}_i^{\text{2-way}}
& \geq q \left( -\tilde{Q}_{\mu_s,i} 
g \left( \tilde{E}_{\mu_s,i} \right)
H_2 \left( \tilde{E}_{\mu_s,i} \right) \right. \nn\\
& \qquad + \left. \sum_{n_i=1}^{2} Q_{n,i}^{L}
\left(1 - G\left(\tilde{e}_{n,i} \right) \right) \right),
\label{eq23}
\end{align}
where the following expressions have been used:
\begin{align}
\tilde{Q}_{\mu_{s, i}} & = Y_0 +\left(1-Y_0\right) \left(1-e^{-\mu_{s,i} T_{i}^{\text{2-way}}}\right) \, , \nn \\
\tilde{E}_{\mu_{s, i}} & = \frac{1}{\tilde{Q}_{\mu_{s, i}}}
\left[e_0 Y_0 + e_{\text{det}}
\left(1-e^{-\mu_{s, i}T_i^{\text{2-way}}}\right) \right] \, ,
\nn \\
G\left(\tilde{e}_{n,i}\right) & =
\begin{cases}
\log_2 \!\left( 1 + 4\tilde{e}_{n,i} - 4\tilde{e}_{n,i}^2 \right), & \tilde{e}_{n,i} < \frac{1}{2} \, ,  \\
1, & \text{otherwise},
\end{cases}
\label{eq24}
\end{align}
where $\tilde{e}_{1,i}$ and $\tilde{e}_{2,i}$ are given in (\ref{eq25}) at the top of the this page.
Following similar steps to the one-way protocol, the corresponding expressions for $Y_{1,i}^{L}$ and $Y_{2,i}^{L}$ are given in (\ref{eq26}) at the top of this page, where the term $Y_{1,i}^{U}$ is derived as:
\setcounter{equation}{26}
\beqarr
Y_{1,i}^{U} = \frac{2\left(Q_{\mu_{1,i}}e^{\mu_{1,i}}-Q_{\mu_{2,i}}e^{\mu_{2,i}}\right)-Y_{2,i}^{\infty}\left(\mu_{1,i}^2 - \mu_{2,i}^2\right)}{2\left(\mu_{1,i}-\mu_{2,i}\right)} \, ,
\label{eq27}
\eeqarr
with $Y_{2,i}^{\infty} = 1-\left(1-Y_0\right)\left(1-T_i^{\text{2-way}}\right)^2$. Moreover, the terms $Q_{n,i}^L$ for $n=1$ and $2$ in (\ref{eq23}) are given as $Q_{1,i}^{L} = Y_{1,i}^{L}e^{-\mu_{s,i}}\mu_{s,i}$ and $Q_{2,i}^{L} = Y_{2,i}^{L}e^{-\mu_{s,i}}\mu_{s,i}^2/2$.
Following this two-way framework, the SKR performance of the MIMO FSO system under the DV-QKD protocol is derived as in (\ref{eq28}) at the top of this page. Moreover, the QBER for single-photon pulses in the considered system setup under the LM05 protocol is calculated as the weighted average of the QBERs $\tilde{e}_{1, i}$ $\forall i$ as follows:
\setcounter{equation}{28}
\beq
\text{QBER}_{\text{MIMO}}^{\text{2-way}} 
= \frac{\sum_{i=1}^{{r_H}} \tilde{e}_{1,i}\,Q_{1,i}^{L}}
{\sum_{i=1}^{{r_H}} Q_{1,i}^{L}} \, .
\label{eq29}
\eeq
%%%%%%%%%%%%%%%%%%%%%%%%%%%%%%%%%%%%%%%%%%%%%%%%%%%%%%%%%%%%%%%%%%%%%%%%%%%%%%%%%%%%%%%%%%%%%%%%%
%%%%%%%%%%%%%%%%%%%%%%%%%%%%%%%%%%%%%%%%%%%%%%%%%%%%%%%%%%%%%%%%%%%%%%%%%%%%%%%%%%%%%%%%%%%%%%%%%
%%%%%%%%%%%%%%%%%%%%%%%%%%%%%%%%%%%%%%%%%%%%%%%%%%%%%%%%%%%%%%%%%%%%%%%%%%%%%%%%%%%%%%%%%%%%%%%%%
\begin{figure*}[!t]
     \centering
     \begin{subfigure}[b]{0.25\textwidth}
         \centering
        \includegraphics[width=4.7cm,height=4cm]{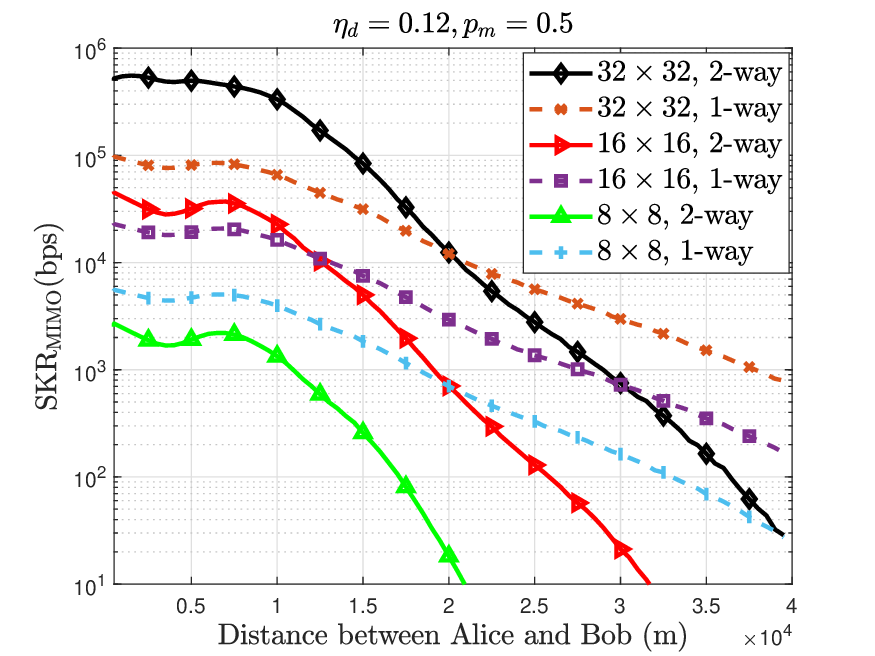}
        \caption{$\text{SKR}_{\text{MIMO}}^{\text{1-way}}$ and $\text{SKR}_{\text{MIMO}}^{\text{2-way}}$ vs. $z$}
         \label{fig;3a}
     \end{subfigure}
     \hspace{-0.3cm}
     \begin{subfigure}[b]{0.25\textwidth}
         \centering
        \includegraphics[width=4.7cm,height=4cm]{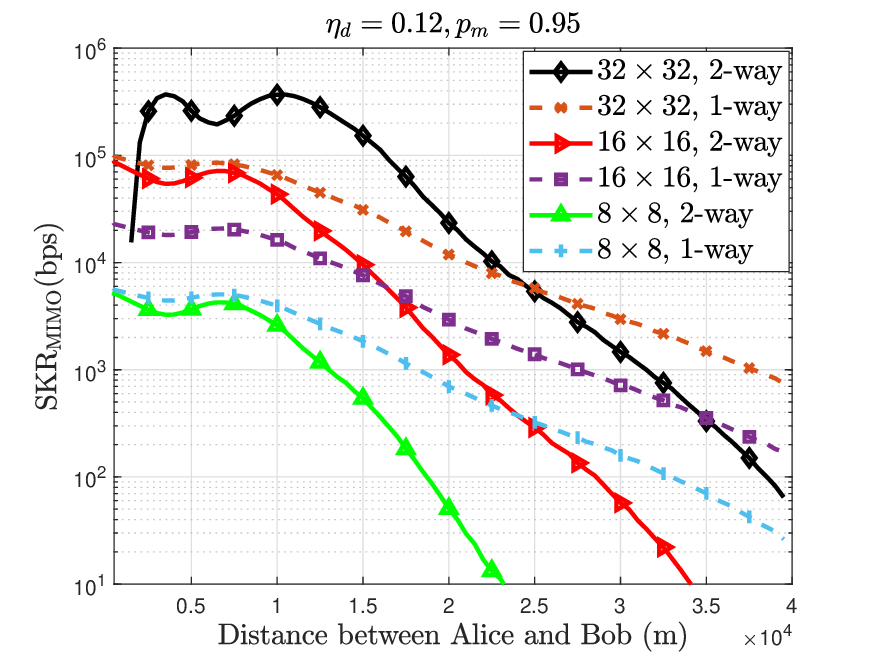}
        \caption{$\text{SKR}_{\text{MIMO}}^{\text{1-way}}$ and $\text{SKR}_{\text{MIMO}}^{\text{2-way}}$ vs. $z$}
         \label{fig;3b}
     \end{subfigure}
     \hspace{-0.3cm}
     \begin{subfigure}[b]{0.25\textwidth}
         \centering
        \includegraphics[width=4.7cm,height=4cm]{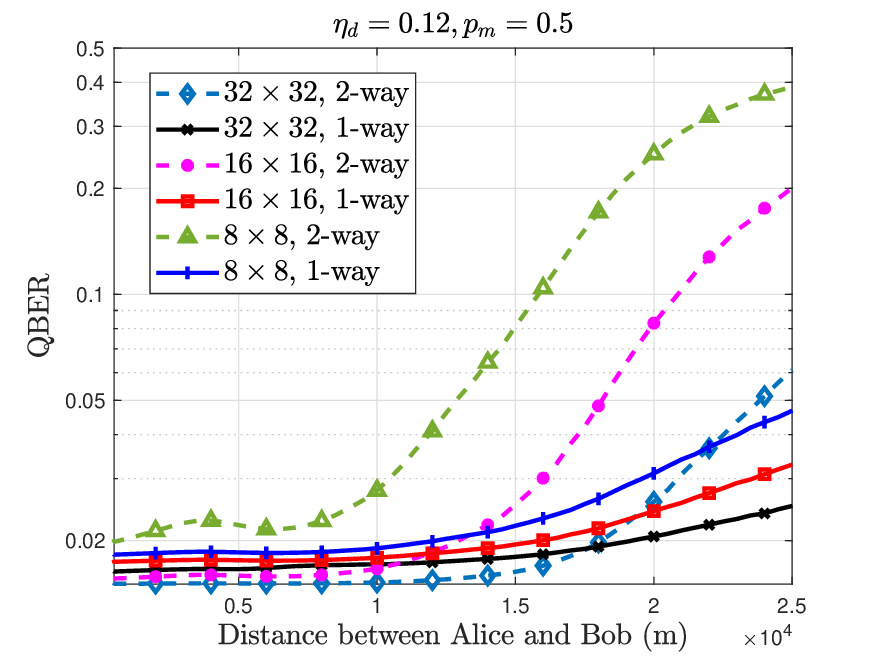}
        \caption{$\text{QBER}_{\text{MIMO}}^{\text{1-way}}$ $\&$ $\text{QBER}_{\text{MIMO}}^{\text{2-way}}$ vs. $z$}
         \label{fig;3c}
     \end{subfigure}
     \hspace{-0.3cm}
     \begin{subfigure}[b]{0.25\textwidth}
         \centering
        \includegraphics[width=4.7cm,height=4cm]{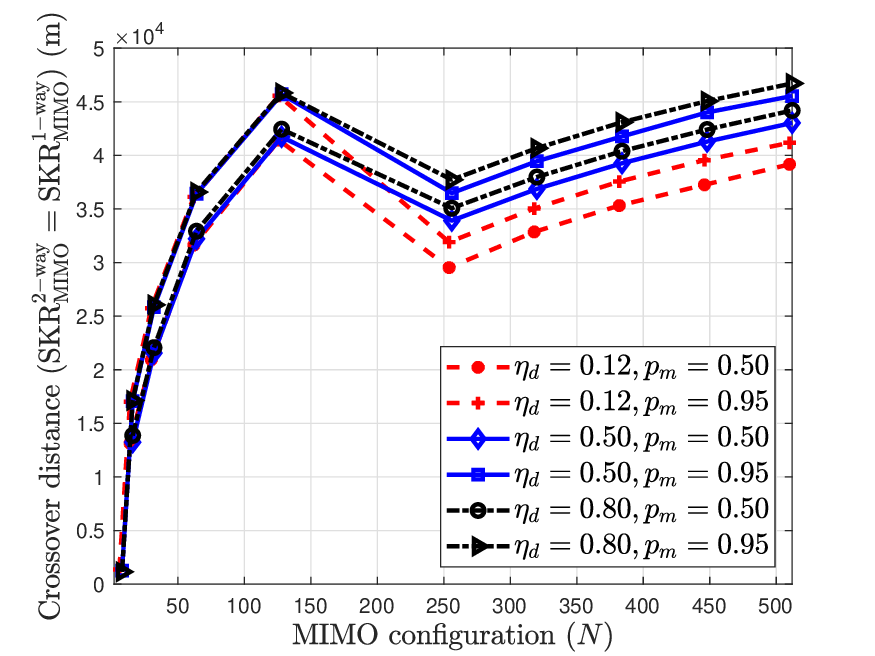}
        \caption{Crossover distance vs. $N$}
         \label{fig;3d}
     \end{subfigure}
    \caption{\small{SKR versus distance between Alice and Bob for the one-way and two-way protocols with $N = 8,16,32$, $\eta_d = 0.12$ and (a) $p_m=0.50$, (b) $p_m=0.95$; (c) QBER versus distance between Alice and Bob for the one-way and two-way protocols with $N=8,16,32$, $\eta_d=0.12$, and $p_m=0.50$; (d) Crossover distance $\left( \text{where } \text{SKR}_{\text{MIMO}}^{\text{2-way}} = \text{SKR}_{\text{MIMO}}^{\text{1-way}}\right)$ versus $N$ for $\eta_d=0.12,0.5,0.8$ and $p_m=0.5,0.95$.}}
     \label{f3}
     \vspace{-0.5cm}
\end{figure*}
%%%%%%%%%%%%%%%%%%%%%%%%%%%%%
\vspace{-0.2cm}
\section{Numerical Results and Discussion}
The system parameters used to generate the numerical results corroborating the presented analysis were: $N_T(=N_R)=N$, $\lambda = 1550 \, \text{nm}$, $ w= 35 \, \text{mm}$ $ a_r= 20 \, \text{cm}$, $\delta=0.43 \times 10^{-3} \, \text{dB/m}$, $C_n =10^{-15}$ $\text{m}^{2/3}$, $\mu_s =0.5$, $\mu_1 = 0.1$, $\mu_2=0.001$, $q =1$ for LM05 protocol, $g\left(E_{\mu_s}\right) = 1.03$, $Y_0 =1.6 \times 10^{-5}$, $e_0 = 0.5$, $e_{\text{det}}=0.015$, and $\theta_p = 1\mu\, \text{rad}$.

Figs.~\ref{fig;3a} and \ref{fig;3b} demonstrate the variations of $\text{SKR}_{\text{MIMO}}$ with the distance between Alice and Bob, considering MIMO configurations with $N = 8, 16$, and $32$ under both the one- and two-way frameworks at $\eta_d = 0.12$ and $p_m =0.50$ as well as at $\eta_d = 0.12$ and $p_m =0.95$, respectively. It is observed that the SKRs decline with increasing distance, while when the MIMO configuration gets larger, the SKR performance improves. Furthermore, the one-way framework exhibits superior performance compared to the two-way one at smaller MIMO settings. On the contrary, the two-way framework outperforms the one-way framework for larger MIMO configurations and at shorter transmission distances. However, this performance gain reverses following a {\em crossover distance} due to several factors, such as round-trip attenuation, cumulative background noise, and detector dark counts in the bidirectional channel. These issues lead to an increased QBER, which further reduces the key rate. It can also be seen that the crossover distance improves with increasing MIMO configuration and the probability of the MM action. Moreover, as expected, the SKR performance also improves at higher $p_m$ values.
% \begin{figure}[!t]
%     \centering
%     \includegraphics[width=8cm,height=6cm]{ICC_skr_dist.eps}
%     \caption{$\text{SKR}_{\text{MIMO}}$ vs. transmission distance for $N=8,16,32$ MIMO configurations for BB84 and LM05 QKD protocol.}
%     \label{f3}
% \end{figure}
% %%%%%%%%%%%%%%%%

Fig.~\ref{fig;3c} illustrates the variation of the QBERs for a single photon of the pulse obtained for the one- and the two-way DV-QKD protocols in \eqref{eq21} and \eqref{eq29}, respectively, with respect to the distance between Alice and Bob for $N = 8, 16$, and $32$ MIMO configurations. It is shown that the QBER gradually increases with distance for all MIMO configurations, attributed to a decrease in optical received power due to path loss and turbulence-induced fading. Notably, the two-way protocol consistently shows a higher QBER compared to the one-way scheme due to the round-trip transmission, which induces additional losses along both the Bob-Alice and Alice-Bob paths. As the number of transmit-receive apertures increases, the overall QBER decreases, which is attributed to the enhanced spatial diversity and the averaging effect across multiple optical paths. However, it is important to note that this averaging effect does not occur in the same proportion for the one- and two-way protocols. Larger MIMO configurations significantly reduce turbulence-induced fading and enhance signal stability, resulting in lower error rates.

Fig.~\ref{fig;3d} depicts the variation of the crossover distance, i.e., the distance at which $\text{SKR}_{\text{MIMO}}^{\text{2-way}}$ = $\text{SKR}_{\text{MIMO}}^{\text{1-way}}$, as a function of the MIMO configuration $N$. It can be seen that, beyond the crossover distance, the one-way protocol starts to outperform the two-way scheme in terms of achievable SKR. It is also observed that, irrespective of the values of $\eta_d$ and $p_m$, the crossover distance tends to increase almost monotonically with $N$. However, the trend reverses for a range of MIMO configurations, following which the crossover distance again increases with increasing $N$ values. Furthermore, no crossover distance is observed for smaller MIMO configurations, which can be attributed to the challenges of compensating losses resulting from lower spatial diversity. As $N$ increases, the system experiences enhanced spatial diversity gain, which allows the two-way scheme to maintain a higher SKR over longer distances. Consequently, the crossover distance gradually increases with the addition of more antennas, demonstrating the robustness of the two-way protocol in large-scale MIMO operations. Moreover, the slight reduction in the crossover distance can be attributed to inter-channel interference, mode misalignment, and hardware limitations, which reduce the theoretical spatial gain that can be achieved with a very large MIMO configuration.
\section{Conclusion}
This paper studied one- and two-way DV-QKD protocols for secret key exchange between two legitimate users connected via MIMO FSO channels. The transmitter LSs used WCPs for secret key generation, the receiver PDs employed threshold detection during the reception of the keys, and an eavesdropper employed a PNS attack to decrypt the keys. Novel expressions for the SKR and QBER performances when employing either of the protocols were derived, which were then corroborated via numerical results. It was observed that, while the inclusion of MIMO improves the performance of the individual protocols consistently over all transmission distances, the two-way protocol outperforms the one-way one only up to a certain transmission distance for larger MIMO configurations.
% Future work will involve the creation of adaptive, hybrid, and experimentally validated frameworks capable of supporting practical quantum networks and key distribution systems integrated with satellites.
%%%%%%%%%%%%%%%%%%%%%%%%%%%%%%%%%%%%%%%%%%%%%%%%%%%%%%%%%%%%%%%%%%%%%%%%%%%%%%%%%%%%%%%%%%%%%%%%%
%%%%%%%%%%%%%%%%%%%%%%%%%%%%%%%%%%%%%%%%%%%%%%%%%%%%%%%%%%%%%%%%%%%%%%%%%%%%%%%%%%%%%%%%%%%%%%%%%
%%%%%%%%%%%%%%%%%%%%%%%%%%%%%%%%%%%%%%%%%%%%%%%%%%%%%%%%%%%%%%%%%%%%%%%%%%%%%%%%%%%%%%%%%%%%%%%%%
%%%%%%%%%%%%%%%%%%%%%%%%%%%%%%%%%%%%%%%%%%%%%%%%%%%%%%%%%%%%%%%%%%%%%%%%%%%%%%%%%%%%%%%%%%%%%%%%%
\bibliographystyle{IEEEtran}
\bibliography{IEEEabrv,bibliography}

% Generated by IEEEtran.bst, version: 1.14 (2015/08/26)
\begin{thebibliography}{10}
\providecommand{\url}[1]{#1}
\csname url@samestyle\endcsname
\providecommand{\newblock}{\relax}
\providecommand{\bibinfo}[2]{#2}
\providecommand{\BIBentrySTDinterwordspacing}{\spaceskip=0pt\relax}
\providecommand{\BIBentryALTinterwordstretchfactor}{4}
\providecommand{\BIBentryALTinterwordspacing}{\spaceskip=\fontdimen2\font plus
\BIBentryALTinterwordstretchfactor\fontdimen3\font minus \fontdimen4\font\relax}
\providecommand{\BIBforeignlanguage}[2]{{%
\expandafter\ifx\csname l@#1\endcsname\relax
\typeout{** WARNING: IEEEtran.bst: No hyphenation pattern has been}%
\typeout{** loaded for the language `#1'. Using the pattern for}%
\typeout{** the default language instead.}%
\else
\language=\csname l@#1\endcsname
\fi
#2}}
\providecommand{\BIBdecl}{\relax}
\BIBdecl

\bibitem{10463684}
M.~A. Jamshed \emph{et~al.}, ``Synergizing airborne non-terrestrial networks and reconfigurable intelligent surfaces-aided {6G IoT},'' \emph{IEEE Internet Things Mag.}, vol.~7, no.~2, pp. 46--52, Mar. 2024.

\bibitem{8920091}
K.~P. Peppas \emph{et~al.}, ``The {F}ischer–{S}nedecor $\mathcal{F}$-distribution model for turbulence-induced fading in free-space optical systems,'' \emph{J. Lightwave Technol.}, vol.~38, no.~6, pp. 1286--1295, 2020.

\bibitem{FSO_secrecy_phuc_sep2020_9194727}
P.~V. Trinh \emph{et~al.}, ``Secrecy analysis of {FSO} systems considering misalignments and eavesdropper’s location,'' \emph{IEEE Trans. Commun.}, vol.~68, no.~12, pp. 7810--7823, Sep. 2020.

\bibitem{bennett1984update}
C.~H. Bennett and G.~Brassard, ``An update on quantum cryptography,'' in \emph{Proc. Workshop Theory Appl. Cryptographic Techniques}.\hskip 1em plus 0.5em minus 0.4em\relax Santa Barbara, California, USA, Aug. 1984, pp. 475--480.

\bibitem{PracticalQKD_Scarani_sep2009}
V.~Scarani \emph{et~al.}, ``The security of practical quantum key distribution,'' \emph{Rev. Mod. Phys.}, vol.~81, pp. 1301--1350, Sep. 2009.

\bibitem{Hwang_aug2003}
W.-Y. Hwang, ``Quantum key distribution with high loss: Toward global secure communication,'' \emph{Phys. Rev. Lett.}, vol.~91, p. 057901, Aug. 2003.

\bibitem{ma_jul2005}
X.~Ma, B.~Qi, Y.~Zhao, and H.-K. Lo, ``Practical decoy state for quantum key distribution,'' \emph{Phys. Rev. A}, vol.~72, p. 012326, Jul. 2005.

\bibitem{Lucamarini_apr2005}
M.~Lucamarini and S.~Mancini, ``Secure deterministic communication without entanglement,'' \emph{Phys. Rev. Lett.}, vol.~94, p. 140501, Apr. 2005.

\bibitem{LUCAMARINI201446}
------, ``Quantum key distribution using a two-way quantum channel,'' \emph{Theoretical Comp. Science}, vol. 560, pp. 46--61, Dec. 2014.

\bibitem{SHAARI2011697}
J.~Shaari, I.~Bahari, and S.~Ali, ``Decoy states and two way quantum key distribution schemes,'' \emph{Opt. Commun.}, vol. 284, no.~2, pp. 697--702, Jan. 2011.

\bibitem{sushil_11129674}
S.~Kumar and S.~P. Dash, ``{SKR} analysis of one- and two-way {CV-QKD MIMO FSO} communication system,'' \emph{IEEE Commun. Lett.}, vol.~29, no.~10, pp. 2456--2460, Oct. 2025.

\bibitem{sushil_RIS}
S.~Kumar, S.~P. Dash, D.~Ghose, and G.~C. Alexandropoulos, ``{RIS}-assisted {MIMO CV-QKD} at {THz} frequencies: Channel estimation and secret key rate analysis,'' \emph{IEEE Trans. Commun.}, early access, 2025.

\bibitem{Zhao_multiplexFSO_dec2015}
N.~Zhao, X.~Li, G.~Li, and J.~M. Kahn, ``Capacity limits of spatially multiplexed free-space communication,'' \emph{Nature Photonics}, vol.~9, no.~12, pp. 822--826, Dec. 2015.

\bibitem{pirandola_FSO_mar2021}
S.~Pirandola, ``Limits and security of free-space quantum communications,'' \emph{Phys. Rev. Res.}, vol.~3, p. 013279, Mar. 2021.

\bibitem{Andrews2005}
L.~C. Andrews and R.~L. Phillips, \emph{Laser Beam Propagation through Random Media}, 2nd~ed.\hskip 1em plus 0.5em minus 0.4em\relax Bellingham, WA: SPIE Press, 2005.

\bibitem{Capraro_turbulance_2012}
{I. Capraro {\em et al.}}, ``Impact of turbulence in long range quantum and classical communications,'' \emph{Phys. Rev. Lett.}, vol. 109, no.~20, p. 200502, Nov. 2012.

\bibitem{gllp_sep2004}
D.~Gottesman, H.-K. Lo, N.~L\"{u}tkenhaus, and J.~Preskill, ``Security of quantum key distribution with imperfect devices,'' \emph{Quantum Inf. Comput.}, vol.~4, no.~5, p. 325–360, Sep. 2004.

\bibitem{beaudry_2013}
N.~J. Beaudry, M.~Lucamarini, S.~Mancini, and R.~Renner, ``Security of two-way quantum key distribution,'' \emph{Phys. Rev. A}, vol.~88, p. 062302, Dec. 2013.

\end{thebibliography}
\end{document}